\documentclass[preprint,12pt]{elsarticle}

\usepackage{amssymb}
\usepackage{float}
\usepackage{xspace}
\usepackage{multirow}
\usepackage{hyperref}
\hypersetup{
    colorlinks,
    linkcolor={red!50!black},
    citecolor={blue!50!black},
    urlcolor={blue!80!black}
}

\usepackage{amsmath}

\newcommand{\nn}{\nonumber}

\newcommand{\br}{{\rm Br}}
\newcommand{\hc}{{\rm H.c.}}

\newcommand{\gev}{{\;{\rm GeV}}}

\newcommand{\beq}{\begin{equation}}
\newcommand{\eeq}{\end{equation}}
\newcommand{\barr}{\begin{array}}
\newcommand{\earr}{\end{array}}
\newcommand{\bc}{\begin{center}}
\newcommand{\ec}{\end{center}}
\newcommand{\bit}{\begin{itemize}}
\newcommand{\eit}{\end{itemize}}
\newcommand{\ben}{\begin{enumerate}}
\newcommand{\een}{\end{enumerate}}

\newcommand{\mh}{m_{h}}
\newcommand{\mch}{M_{H^\pm}}
\newcommand{\mhh}{M_{H}}
\newcommand{\ma}{M_{A}}

\newcommand{\tb}{\tan\beta}

\newcommand{\ttop}      {{t\bar{t}}}

\journal{Nuclear Physics B}

\begin{document}

\begin{frontmatter}



\title{Can a pseudoscalar with a mass of 365 GeV in the 2HDM explain the CMS $t\bar{t}$ excess?}


\author[label1]{Chih-Ting Lu} 
\ead{ctlu@njnu.edu.cn}

\author[label2,label3]{Kingman Cheung } 
\ead{cheung@phys.nthu.edu.tw}

\author[label4]{Dongjoo Kim}
\ead{dongjookim.phys@gmail.com}

\author[label2]{Soojin Lee}
\ead{soojin.lee@gapp.nthu.edu.tw}

\author[label4]{Jeonghyeon Song}
\ead{jhsong@konkuk.ac.kr}

\affiliation[label1]{organization={Department of Physics and Institute of Theoretical Physics, Nanjing Normal University},
            city={Nanjing},
            country={China}}
\affiliation[label2]{organization={Department of Physics, National Tsing Hwa University},
            city={Hsinchu},
            country={Taiwan}}
            
            \affiliation[label3]{organization={Center for Theory and Computation, National Tsing Hua University},
             city={Hsinchu},
            country={Taiwan}}

 \affiliation[label4]{organization={Department of Physics, Konkuk University},
             city={Seoul},
            country={Republic of Korea}}

\begin{abstract}
We analyze the CMS-reported $t\bar{t}$ excess within conventional Two-Higgs-Doublet Models (2HDMs) of Types I, II, X, and Y, using the best-fit pseudoscalar parameters $M_A = 365$ GeV, $\Gamma_A/M_A = 2\%$, and $\tan\beta = 1.28$. Applying theoretical and experimental constraints—including stability, unitarity, perturbativity, FCNCs, and collider bounds—we find perturbativity limits $M_{H^\pm}$ and $M_H$ to below about 723 GeV. FCNC constraints exclude Types II and Y, while the remaining parameter space in Types I and X is ruled out by recent $t\bar{t}Z$ measurements from ATLAS and CMS. We conclude that conventional 2HDMs cannot explain the observed $t\bar{t}$ excess, though toponium effects in the background fit may change this conclusion.\end{abstract}


\end{frontmatter}

\section{Introduction}
\label{sec1}

Understanding the fundamental constituents of matter and their interactions remains a central goal of particle physics. 
The Standard Model (SM) has been remarkably successful, yet its domain of validity must be tested with ever-increasing precision; in this context, the Large Hadron Collider (LHC) plays a dual role by performing precision measurements and probing for signs of physics beyond the SM.
 As a prolific top-quark factory, the LHC enables both high-precision measurements of top-quark properties and sensitive searches for new resonances in $t\bar t$ final states. In particular, production near the $t\bar t$ threshold provides a unique window into new dynamics, although it is intrinsically challenging due to non-perturbative QCD effects~\cite{Sumino:1992ai,Hagiwara:2008df}.

Recently, the CMS Collaboration reported an excess in the $t\bar t$ invariant-mass spectrum near threshold at $\sqrt{s}=13$ TeV using 138 fb$^{-1}$ of data~\cite{CMS-PAS-HIG-22-013}. The best-fit interpretation, with a significance exceeding $5\sigma$, corresponds to a spin-singlet toponium state $\eta_t$~\cite{Fuks:2021xje}. An alternative explanation in terms of a fundamental pseudoscalar boson $A$ with $M_A=365$ GeV and $g_{Att}=0.78$ also provides a good fit and is strongly favored by CP-sensitive observables over a scalar hypothesis. This raises the important question of whether such a pseudoscalar can be consistently embedded in a well-motivated ultraviolet-complete framework.

Motivated by this observation, we investigate the pseudoscalar interpretation within conventional Two-Higgs-Doublet Models (2HDMs)~\cite{Branco:2011iw}, which naturally contain a CP-odd scalar and were used as a prototype in the CMS analysis~\cite{CMS-PAS-HIG-22-013}. We examine the viability of the relevant parameter space under theoretical consistency conditions~\cite{Eriksson:2009ws}, flavor  constraints~\cite{Misiak:2017bgg}, electroweak precision data, collider searches, and in particular the latest $t\bar t Z$ measurements from ATLAS and CMS~\cite{ATLAS:2023szc,CMS:2024mtn}.

\section{CMS $t\bar{t}$ Excess in the 2HDM}
\label{sec:review}

The 2HDM  extends the Standard Model by introducing two complex $SU(2)_L$ Higgs doublets, $\Phi_1$ and $\Phi_2$~\cite{Branco:2011iw},
which have non-zero vacuum expectation values (VEVs), $v_1$ and $v_2$, respectively. Electroweak symmetry breaking occurs at
$v=\sqrt{v_1^2+v_2^2}=246\gev$, and we define $\tan\beta=v_2/v_1$.

To forbid tree-level Flavor-Changing Neutral Currents (FCNCs),
we impose a discrete $Z_2$ symmetry under which $\Phi_1\to\Phi_1$ and $\Phi_2\to-\Phi_2$.
The $CP$-conserving scalar potential with a softly broken $Z_2$ symmetry is
\begin{align}
\label{eq-VH}
V ={}& m_{11}^2\,\Phi_1^\dagger\Phi_1
+ m_{22}^2\,\Phi_2^\dagger\Phi_2
- m_{12}^2\left(\Phi_1^\dagger\Phi_2+\hc\right) \nn\\
&+ \frac{1}{2}\lambda_1(\Phi_1^\dagger\Phi_1)^2
+ \frac{1}{2}\lambda_2(\Phi_2^\dagger\Phi_2)^2
+ \lambda_3(\Phi_1^\dagger\Phi_1)(\Phi_2^\dagger\Phi_2)
+ \lambda_4(\Phi_1^\dagger\Phi_2)(\Phi_2^\dagger\Phi_1) \nn\\
&+ \frac{1}{2}\lambda_5\left[(\Phi_1^\dagger\Phi_2)^2+\hc\right].
\end{align}

After electroweak symmetry breaking, the model contains five physical Higgs bosons:
two $CP$-even scalars $h$ and $H$, a $CP$-odd pseudoscalar $A$, and a charged Higgs pair
$H^\pm$. The weak eigenstates are related to the mass eigenstates through the mixing
angles $\alpha$ and $\beta$. The SM-like Higgs boson is given by
\begin{equation}
h_{\rm SM} = \sin(\beta-\alpha)\,h + \cos(\beta-\alpha)\,H,
\end{equation}
and throughout this work we identify $h$ with the observed Higgs boson,
$\mh=125\gev$.

The Yukawa interactions depend on the $Z_2$ charge assignments of the fermion singlets,
giving rise to four canonical realizations: Type-I, Type-II, Type-X, and Type-Y.
Our analysis shows that Type-I (Type-II) yields phenomenologically identical results
to Type-X (Type-Y) for the CMS $\ttop$ excess. We therefore focus on Type-I and Type-II.

In the alignment limit, the Yukawa coupling modifiers are
\begin{align}
\text{Type-I:}\quad
& \xi_u^H=\xi_d^H=\xi_\ell^H
= \xi_u^A=-\xi_d^A=-\xi_\ell^A
= \frac{1}{\tan\beta},
\\ \nn
\text{Type-II:}\quad
& \xi_u^H=\xi_u^A=\frac{1}{\tan\beta}, \qquad
\xi_d^H=\xi_\ell^H=\xi_d^A=\xi_\ell^A=\tan\beta .
\end{align}

The CMS Collaboration, using 138 fb$^{-1}$ of data at the 13 TeV LHC, reported an excess in the
$t\bar t$ invariant-mass spectrum around 365 GeV. Among the interpretations considered,
a $^1S_0^{(1)}$ toponium state $\eta_t$ provides the best fit, while a single pseudoscalar
$A$ with $m_A=365$ GeV also yields a local significance above $5\sigma$ and is strongly
favored over a scalar hypothesis by CP-sensitive observables. This indicates that a
fundamental pseudoscalar offers an explanation of the excess that is competitive with the
toponium scenario, arising from gluon-fusion production followed by $A\to t\bar t$.

Motivated by the CMS best-fit results, we adopt the following benchmark setup:~\cite{Lu:2024twj}
\begin{align}
\label{eq-setup}
\mh &= 125\gev, \qquad \sin(\beta-\alpha)=1,
\\ \nn
\ma &= 365\gev, \qquad \tan\beta = 1.28 \pm 0.128 ,
\end{align}
allowing for a 10\% uncertainty in $\tan\beta$.

To assess whether the pseudoscalar $A$ can simultaneously satisfy
$\br(A\to t\bar t)\simeq1$ and $\Gamma_A/m_A\simeq2\%$, we study its decay properties
as a function of the charged Higgs mass, assuming $m_{H^\pm}=m_H$.
For $m_A=365\gev$ and $\tan\beta=1.28$, we find no qualitative difference between
Type-I and Type-II.

The decay pattern of $A$ is governed by the masses of $H$ and $H^\pm$.
When these states are light, the channels $A\to H^\pm W^\mp$ and $A\to ZH$ open and
substantially reduce $\br(A\to t\bar t)$. Once $m_{H^\pm}=m_H\gtrsim280\gev$, these
modes are kinematically suppressed and $\br(A\to t\bar t)\simeq1$ is achieved.
The total width follows a similar trend: for lighter $H$ and $H^\pm$,
$\Gamma_A/m_A$ exceeds the CMS-preferred value, while heavier masses yield
$\Gamma_A/m_A\simeq1.8\%$, in excellent agreement with the best fit.
This implies that explaining the CMS excess requires
$m_{H^\pm}\simeq m_H \gtrsim 280\gev$.

We next examine the viability of these parameter points under theoretical and
experimental constraints by performing a random scan over
\begin{align}
\label{eq-scan-range}
\tan\beta &\in [1.152,1.408], & m_{12}^2 &\in [0,1000^2]\gev^2,
\\ \nn
m_H &\in [130,1500]\gev, & m_{H^\pm} &\in [200,1500]\gev,
\end{align}
assuming the setup in \autoref{eq-setup}.

\begin{figure}[t]
\centering
\includegraphics[width=0.95\textwidth]{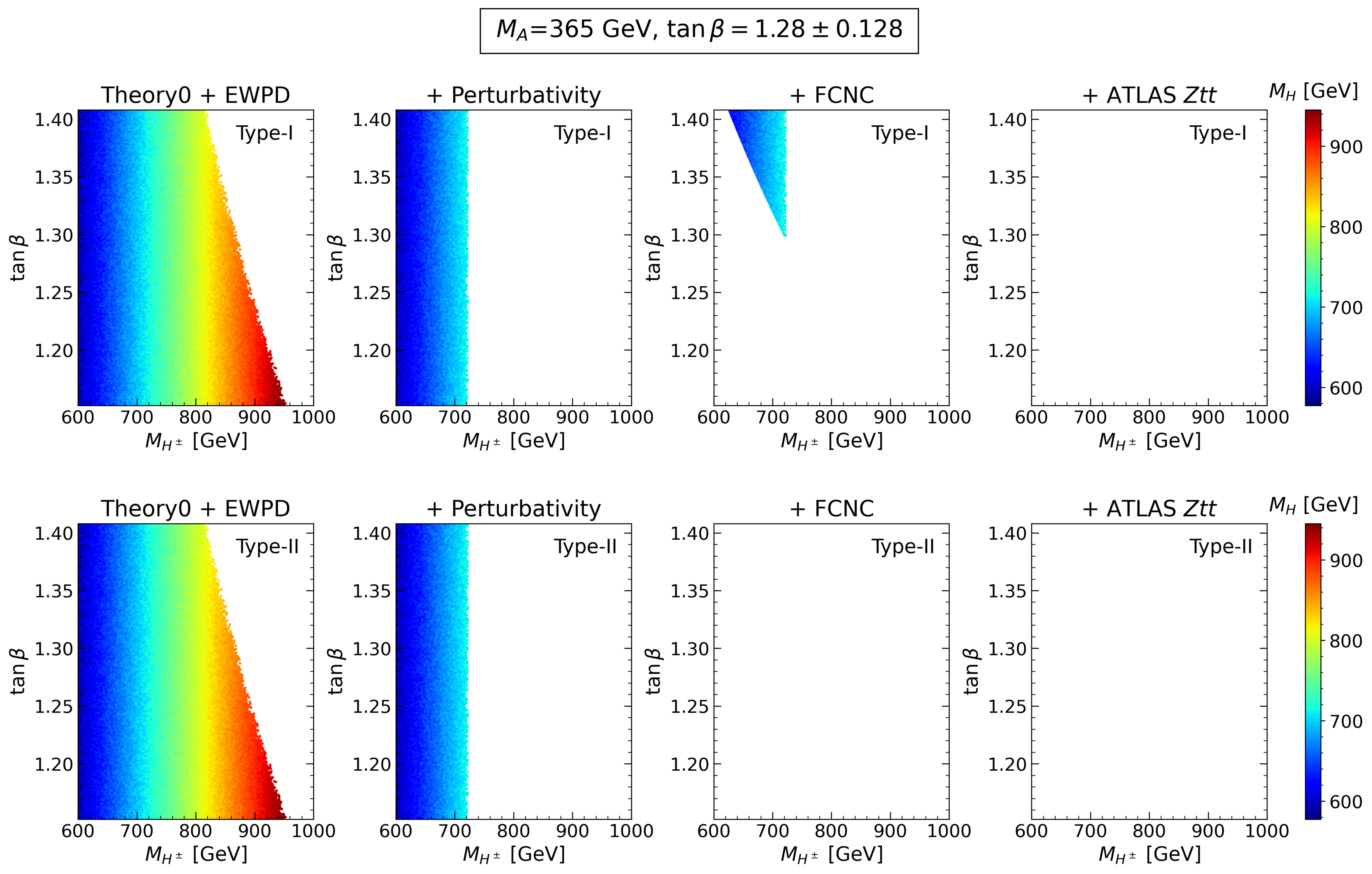}
\caption{\label{fig-survived-param}
Surviving parameter space in the $(M_{H^\pm},\tan\beta)$ plane with $M_H$ shown by color.
Upper (lower) panels correspond to Type-I (Type-II).
Columns show cumulative constraints from Steps (i) to (iv).
All panels assume $M_A=365\gev$ and $\tan\beta=1.28\pm0.128$.}
\end{figure}

We impose constraints cumulatively:
(i) bounded-from-below potential, unitarity, vacuum stability, and oblique parameters;
(ii) perturbativity, requiring $|C_{H_iH_jH_kH_l}|<4\pi$;
(iii) FCNC bounds from $b\to s\gamma$ and $B_d\to\mu\mu$~\cite{Misiak:2017bgg};
(iv) Higgs precision data and direct collider searches using
\textsc{HiggsTools}.
Steps (i) and (ii) are evaluated with \textsc{2HDMC} v1.8.0~\cite{Eriksson:2010zzb}.
The results are shown in \autoref{fig-survived-param}.

After Step (i), unitarity restricts $m_{H^\pm},m_H\lesssim950\gev$, and oblique
parameters enforce a strong correlation $m_{H^\pm}\simeq m_H$, reflecting
\[
\lambda_4-\lambda_5 = 2\frac{m_A^2-m_{H^\pm}^2}{v^2}.
\]
Perturbativity (Step ii) further excludes all points with
$m_{H^\pm}\simeq m_H\gtrsim723\gev$. These violations dominantly affect Higgs
self-couplings involving the SM-like Higgs, leading to unacceptable loop-level
deviations, and therefore cannot be relaxed.

Flavor constraints (Step iii) leave only a small region with
$\tan\beta\gtrsim1.3$ in Type-I, while Type-II is entirely excluded due to the
incompatibility between $b\to s\gamma$ bounds and perturbativity.
Finally, collider constraints (Step iv) eliminate the remaining Type-I points
through $pp\to Zt\bar t$ searches at the LHC~\cite{ATLAS:2023szc,CMS:2024mtn}.
Although designed for $H\to ZA$, these searches also constrain
$pp\to H\to ZA\to Zt\bar t$, which applies here given the sizable gluon-fusion
rate for $m_H\lesssim723\gev$.

\begin{figure}[t]
\centering
\includegraphics[width=0.85\textwidth]{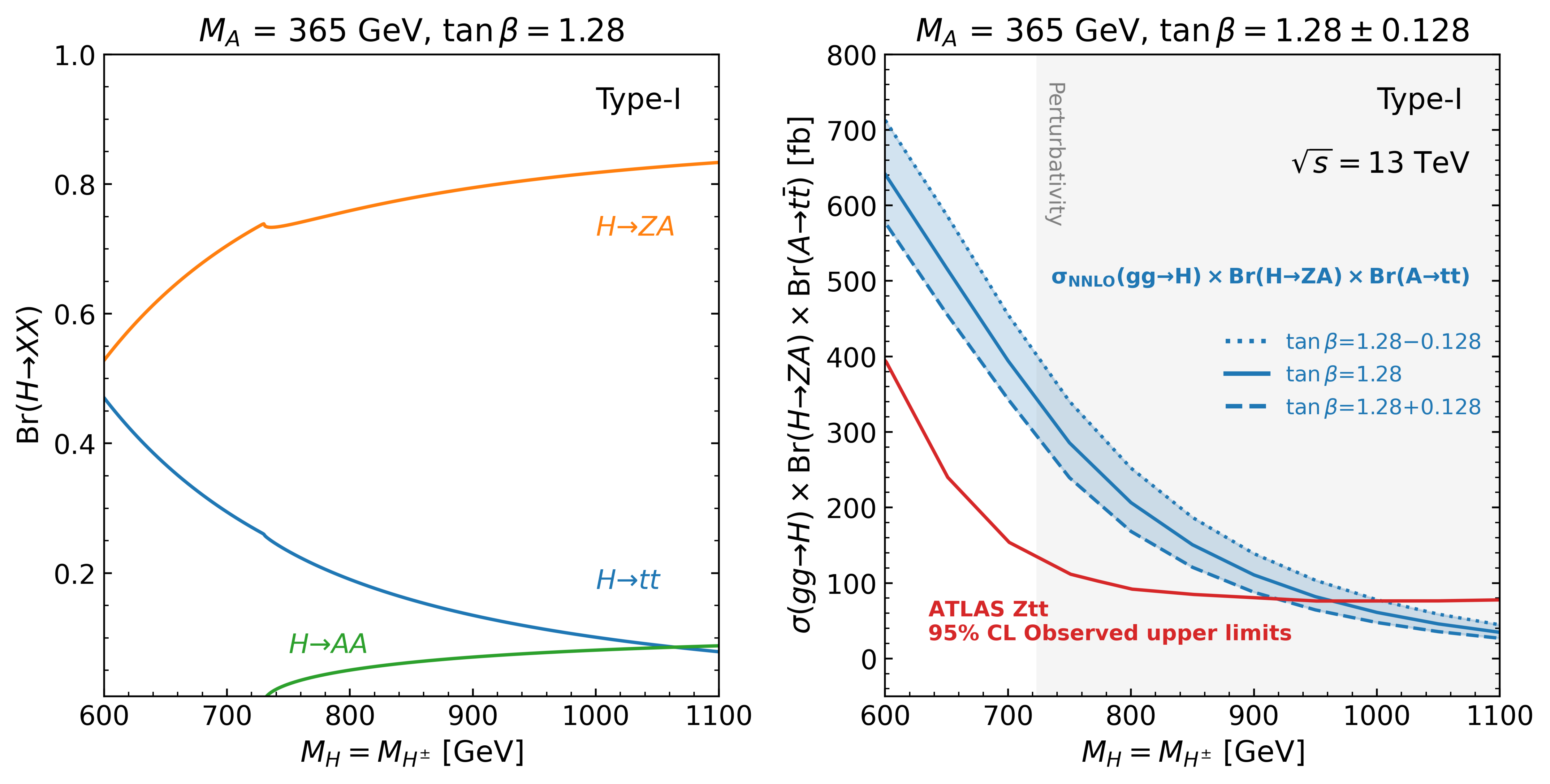}
\caption{\label{fig-BRH-Ztt}
Left: $\br(H\to ZA)$ as a function of $\mhh$.
Right: $\sigma_{\rm NNLO}(gg\to H)\br(H\to ZA)\br(A\to\ttop)$ at 13 TeV.
Parameters are $\ma=365\gev$, $\tb=1.28$, $m_{12}^2=7\times10^4\gev^2$, and $\mch=\mhh$.}
\end{figure}

As shown in \autoref{fig-BRH-Ztt}, the predicted $Zt\bar t$ rate exceeds the
experimental bound unless $m_H\gtrsim900\gev$, a region already excluded by
perturbativity. We therefore conclude that conventional 2HDMs cannot accommodate
the CMS $\ttop$ excess.

\section{Conclusions}
\label{sec:conclusions}

The CMS Collaboration has reported an excess in $t\bar t$ production near an invariant mass of 365 GeV, with angular observables pointing to a pseudoscalar origin. We investigated whether this excess can be explained by the pseudoscalar boson of conventional Two-Higgs-Doublet Models (2HDMs).

Using the CMS best-fit values $m_A=365$ GeV, $\Gamma_A/m_A=0.02$, and $\tan\beta=1.28$, allowing a 10\% variation in $\tan\beta$, we performed a comprehensive scan of Type-I and Type-II 2HDMs, effectively covering all four canonical types. Imposing theoretical consistency, electroweak precision data, perturbativity, flavor constraints, and the latest $t\bar t Z$ limits, we find that perturbativity restricts additional Higgs masses to $\lesssim723$ GeV. Flavor constraints exclude all of Type-II, while the small surviving region in Type-I is ruled out by recent ATLAS $t\bar t Z$ searches.

We therefore conclude that conventional 2HDMs (Types I, II, X, and Y) cannot accommodate a pseudoscalar explanation of the CMS $t\bar t$ excess. More general extensions of the 2HDM may still be viable. We also note that including toponium effects in the background modeling could reopen limited parameter space at larger $\tan\beta$.

 \bibliographystyle{elsarticle-num} 
 \bibliography{Song-TOP2025}






\end{document}